\documentclass[9pt,twocolumn,twoside]{opticajnl}
\journal{opticajournal} 
\setboolean{shortarticle}{true}
\usepackage{lineno}
\title{Fully guided and phase locked Ti:PPLN waveguide squeezing for applications in quantum sensing}

\author[1,*]{Renato Domeneguetti}
\author[2]{Michael Stefszky}
\author[2]{Harald Herrmann}
\author[2]{Christine Silberhorn}
\author[1]{Ulrik L. Andersen}
\author[1]{Jonas S. Neergaard-Nielsen}
\author[1,$\dagger$]{Tobias Gehring}

\affil[1]{Center for Macroscopic Quantum States bigQ, Department of Physics, Technical University of Denmark, Fysikvej 307, DK-2800 Kgs.Lyngby, Denmark.}
\affil[2]{Integrated Quantum Optics Group,
Institute of Photonic Quantum Systems (PHOQS),
University Paderborn,
Warburger Str. 100,
33098 Paderborn,
Germany}

\affil[*]{Corresponding author: rerdo@dtu.dk}
\affil[$\dagger$]{Corresponding author: tobias.gehring@fysik.dtu.dk}

\begin{abstract} 
This work reports a fully guided setup for single-mode squeezing generation on integrated titanium-indiffused periodically poled nonlinear resonators. A continuous wave laser beam is delivered and the squeezed field is collected by single-mode fibers, where up to -3.17(9)~dB of useful squeezing is available in fibers. To showcase the usefulness of such a fiber-coupled device, we applied the generated squeezed light in a fiber-based phase sensing experiment, showing a quantum enhancement in the signal-to-noise ratio of 0.35~dB. Moreover, our investigation of photorefraction's impact on the cavity resonance condition suggests that it causes system instabilities at high powers.
\end{abstract}

\setboolean{displaycopyright}{false}

\begin{document}

\maketitle

\paragraph{Introduction.} Squeezed states of light are an essential tool for enhancing the sensitivity of measurements of physical quantities with the help of phase or amplitude quadratures of an electromagnetic field~\cite{andersen_30_2016}. Often interferometers are used for those measurements, and by replacing the vacuum with squeezed states in the dark port of the main beam splitter, the shot-noise limit (SNL) can be overcome and the signal-to-noise ratio (SNR) is enhanced by an amount commensurate to the squeezing level~\cite{caves_quantum-mechanical_1981,xiao_physical_1987}. For example, squeezed states are successfully used to enhance the sensitivity of gravitational-wave detectors~\cite{abadie_gravitational_2011,aasi_enhanced_2013}.

A squeezed light source that is simultaneously scalable, reproducible, stable, and compact is appealing since it will allow the mass fabrication of consistent and robust photonic chips. Moreover, for squeezed light generation in the telecommunication C-band, it is interesting to deliver the squeezed light in fiber, with the potential of integrating the device into an already existing optical setup or fiber network. A fully guided squeezing source was reported that combined a commercially available spontaneous parametric down-conversion (SPDC) module with other optical fiber components using simple plug-and-play assembly, but with almost no room for improvement of its quantum efficiency~\cite{kaiser_fully_2016}.

Various on-chip nonclassical light sources have recently been reported due to technological advances in fabricating photonic devices with low propagation loss waveguides. For example, broadband and up to 6.3 dB of squeezed light was achieved in a periodically poled lithium-niobate (PPLN) ridge waveguide in a single-pass configuration thanks to the high second-order nonlinearity of the lithium-niobate crystal, but at the cost of high pump power in the absence of a cavity. The output was collected in free space~\cite{kashiwazaki_continuous-wave_2020} and with a fiber~\cite{kashiwazaki_fabrication_2021}. A reconfigurable chip for the generation of single- or two-mode squeezing in PPLN working at high pump power was described in~\cite{lenzini_integrated_2018}. A two-module PPLN device containing cascaded second-harmonic generation (SHG) and optical-parametric amplifier (OPA) produced 1\,dB of squeezing in free space~\cite{Kanter:02}.

Several integrated sources of single- or two-mode squeezed light using silicon photonics have also been reported \cite{dutt_-chip_2015,zhao_near-degenerate_2020,vaidya_broadband_2020,zhang_squeezed_2021}. Silicon nitride chips have a low third-order nonlinearity, offering weak light-matter interactions, which can be compensated for with high-quality-factor resonators, made possible by silicon nitride's low propagation losses and high confinement of light. Thus, high on-chip levels of squeezing were achieved, but only small levels were actually measured due to the difficulty of efficiently coupling light from chip to single-mode fibers.

Here, we report broadband squeezing generation in a titanium indiffused periodically poled lithium niobate (Ti:PPLN) waveguide resonator for quantum sensing applications. In Ref.~\cite{stefszky_waveguide_2017}, squeezing was demonstrated with these geometries in a free space setup, where a total detection efficiency, measured from the output of the chip to the homodyne detector, of 72\,\% was achieved. In that work, to reach high homodyne detection visibility, a second waveguide was used as a mode shaper for the local oscillator (LO). We instead constructed a fully guided setup, where the squeezed light was directly coupled into a single-mode fiber and the pump field was coupled from fiber to the chip. Homodyne detection was performed using a fiber coupler with high fringe visibility. Despite the fiber components, we achieved a total detection efficiency comparable to the free space setup, with the advantage of being more compact, robust, and ready to be integrated into different fiber-coupled setups. We furthermore locked many degrees of freedom in our setup: fiber alignment, pump-to-seed phase, and the LO phase of homodyne detection. Finally, we demonstrated the applicability of our squeezer in a quantum sensing experiment, where vacuum fluctuations were replaced by phase squeezing to enhance the signal-to-noise ratio in a phase sensing protocol.

\paragraph{Squeezed Light Source.} The central part of the squeezed light source was an optical waveguide cavity in Ti:PPLN. In the cavity, squeezed light was generated through a $\chi^{(2)}$-nonlinear interaction and quasi-phase matching was achieved via the periodic inversion of the ferroelectric domains with a proper poling period. Optical waveguides in the 12~mm long PPLN crystal were fabricated via indiffusion of Ti-stripes into the material, a mature and well-known fabrication technology to obtain low-loss waveguides~\cite{Santandrea_2019}. The waveguide fabrication conditions were chosen to enable single-mode guiding in the telecommunication C-band. For such optimized waveguides, the losses are about 10\,dB/m, with the best ones at around 2\,dB/m, mainly caused by scattering in waveguide surfaces~\cite{luo_direct_2015}.

For the nonlinear conversion process, we exploited the strongest nonlinear coefficient $\mathrm{d_{33}}$ of lithium niobate, which requires quasi-phase-matching with poling periods around 17 µm for a type 0 process, where all the interacting waves are extraordinarily polarized. Operation at elevated temperature $\mathrm{\sim140.5^{\circ}C}$ is necessary to mitigate the photorefractive effect~\cite{rams_optical_2000}.

To reduce the pump power required for strong nonlinear conversion, dielectric mirrors were deposited on the waveguide end facets to form a cavity. The reflectivities of these mirrors were optimized on the basis of numerical simulation results for maximum squeezing and low power threshold. The input mirror comprised a SiO2/TiO2 dielectric stack with 15 layers, which yielded a high reflectivity (HR) (>99~\%) in the telecom C-band. The 11-layer stack of the output mirror had a partial reflectivity (PR) of about 64~\%. The transmission of both mirrors at the pump wavelength was higher than 97~\%. No cavity for the pump is allowed since it can increase the photorefraction~\cite{mondain_photorefractive_2020}.

\begin{figure}[htbp]
\centering
\includegraphics[width=.45\textwidth]{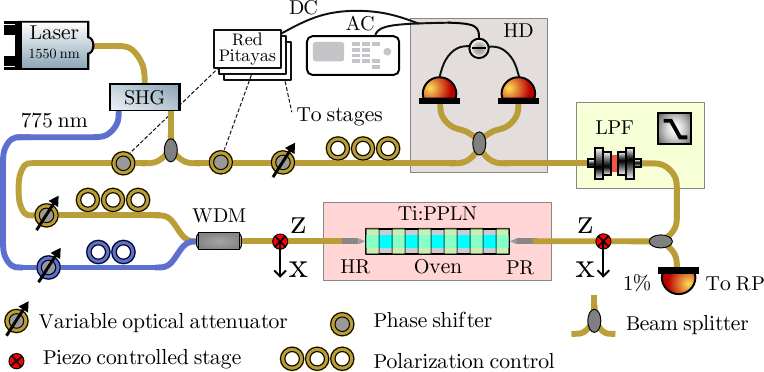}
\caption{\label{fig:setup} Experimental setup for the quantum characterization and operation of the Ti:PPLN squeezer. SHG, second harmonic generation (frequency doubler); WDM, wavelength-division multiplexer, HR-PR high-partial reflecting coatings; LPF low-pass filter; HD homodyne detection; DC-AC low-high electronic frequencies components; To RP means to a Red Pitaya.}
\end{figure}

The setup for quantum characterization and operation of the Ti:PPLN squeezer is shown in Fig.~\ref{fig:setup}. Light at 1550~nm from a  continuous-wave (CW) fiber laser was frequency doubled to 775~nm (NKT Koheras HARMONIK). Part of the non-converted light at 1550~nm was used as a probe for phase locking the OPO to produce squeezing at a specific quadrature, and as a local oscillator for homodyne detection. Both pump and probe fields were power and polarization controlled before coupling them into the nonlinear waveguide by using variable optical attenuators and polarization controllers in single-mode (SM) fiber at respective bands. Then, both light fields were combined in one SM fiber at 1550~nm via a wavelength-division multiplexer (WDM) from Thorlabs (WD1350A). Although this WDM is designed to operate at 1310/1550~nm, it worked very well for the pump field, showing an insertion loss of only 1~dB. Also, the output of the pump field out of the SM fiber was always transversally fundamental regardless of the power, polarization, or  curvature of the fiber. The coupling of both fields into the nonlinear resonator was facilitated by a lensed fiber with 4~µm spot size which according to simulations matches better with the mode-field diameter of the waveguide for the pump field. Furthermore, the lensed fiber has a finite working distance, allowing maximum coupling efficiency before touching the waveguide facet. The maximum coupling efficiency of the pump beam into the waveguide was estimated to be 70\% by measuring the power at the fiber output and comparing it with the power at the output of the waveguide on the other side after discounting propagation losses at 775\,nm. 

A lensed fiber with 6~µm spot size collected the squeezed vacuum out of the chip at the PR port of the resonator, where a maximum coupling efficiency of 88\% was reached. The estimation was accomplished by comparing the measured output power of the probe with an aspherical anti-reflection coated lens, followed by measuring the power collected via lensed fiber. Because of the cavity behavior, both measurements were realized with the cavity in resonance with the probe. It is important to remark that a coupling efficiency of 90~\% was achieved using a butt-coupled AR-coated fiber instead of lensed fiber. However, we preferred to work with lensed fiber because of its finite working distance, making the alignment easier when no visual access to the coupling is available. Nevertheless, pigtailing is a good option for commercial devices when no further characterization is needed.

The collected squeezing was guided to homodyne detection via SM fiber. Here, 1\% of the co-propagated probe was tapped off for locking the squeezing phase and to perform self-alignment of fibers. Next, it passed through a low-pass filter (LPF) from Omega Optical LLC to eliminate the pump field from the fiber. The intrinsic loss of the LPF was less than 2\,\% as indicated in the datasheet, but 8\,\% of loss was measured which can be attributed to the mating sleeve in this connection. Homodyne detection was performed using a 3\,dB coupler and a custom-made homodyne photodetection circuit.

Apart from the connection to the LPF we spliced all fiber connections to minimize loss. Therefore, we achieved an estimated total propagation efficiency from the output of the resonator until the input port of the beam splitter (BS) for homodyne detection of 80~\%. Because both the squeezing and LO propagates in SM fiber, the visibility of the homodyne detection is very high, ~$\sim$~99\,\%. Additionally, the quantum efficiency of the homodyne detection was estimated to be 93.4~$\pm$~1.0~\%. Thus, the total detection efficiency was estimated to be~$\sim$~74\,\%.\

\begin{figure}[htbp]
\centering
\includegraphics[width=.45\textwidth]{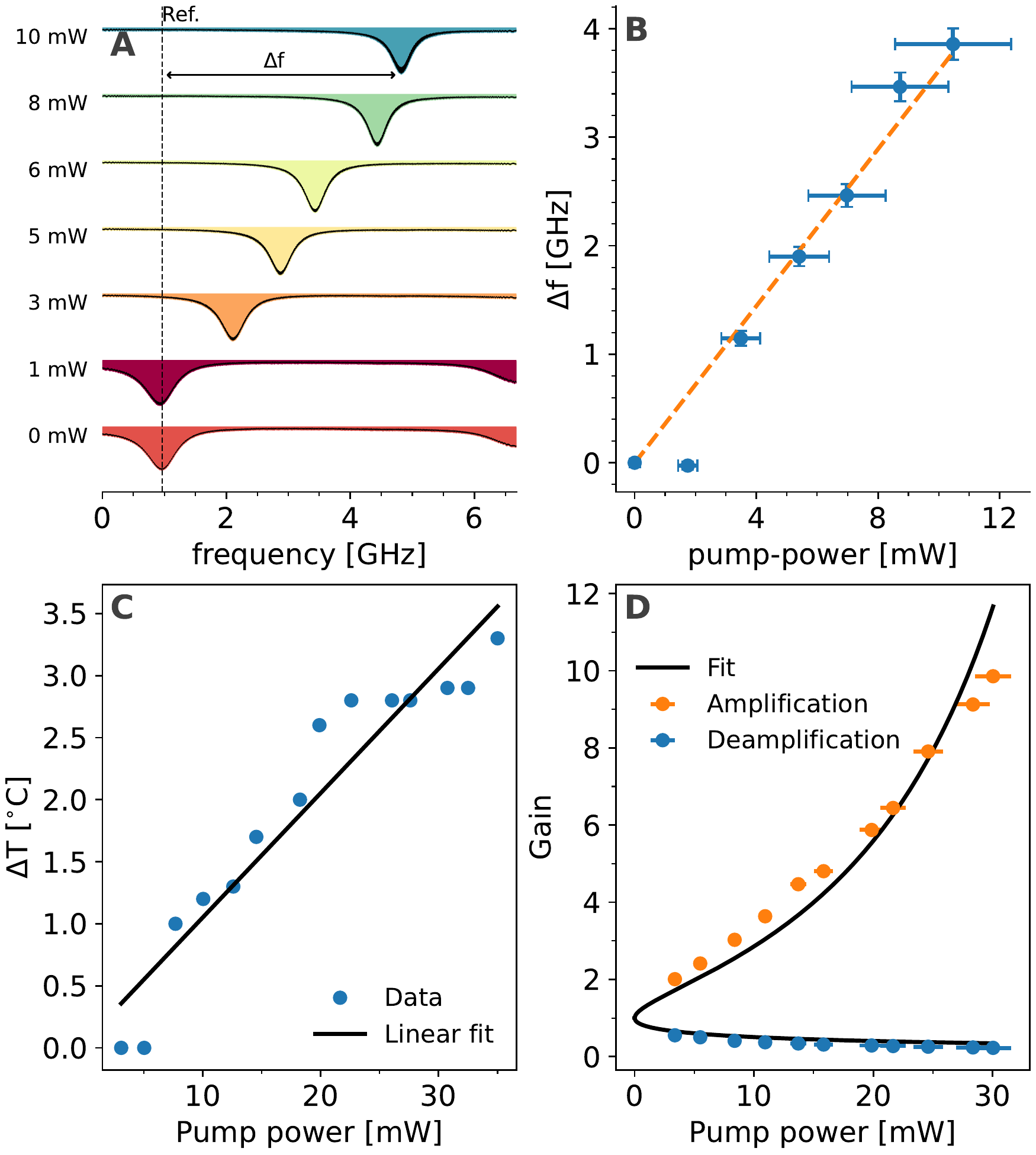} 
\caption{\label{fig:classical} Classical characterization of the Ti:PPLN squeezer. A) Blueshift was observed when increasing pump power. B) Estimation of the photorefraction effect. C) Increase in the phase-matching temperature due to the photorefraction. D) PDC gain after correction.}
\end{figure}

\paragraph{Locking system.} To enable locking of the system, the probe beam was phase modulated at 60\,kHz which was detected after the cavity at the 1\,\% port of the beam splitter. A Red Pitaya (STEM 125-14) generated the error signal by demodulating the electronic signal and produced a control signal for the actuation of the phase shifter to achieve deamplification. To avoid noise contamination from the probe beam to the squeezing, the maximum power of the probe before the HR port of the resonator had to be less than 10~µW as only in this case the probe was shot noise limited at analysis frequencies above 10~MHz. For the phase of the LO, the LO beam was phase modulated at 55\,kHz which was detected by the homodyne detector feeding another Red Pitaya. We furthermore implemented a self-alignment algorithm based on gradient descent to improve the stability of the alignment between the input/output fiber and the waveguide, which suffered from small but continuous piezo drifts that limit the optimum alignment to a couple of seconds. Although all locks worked well for low pump powers (< 10 mW of power inside the waveguide), they failed to maintain the experiment stability when increasing the input power, which we mainly attributed to the photorefractive process~\cite{savchenkov_enhancement_2006}. At high pump powers, small fluctuations in the coupling efficiency of light from fiber to waveguide can notably change the power in the waveguide, where both thermal effects and photorefraction alter the resonance condition, rendering the lock ineffective. In fact, if the power is high enough, the system is expected to enter a chaotic regime due to photorefraction~\cite{carrascosa_understanding_2008}.

\paragraph{Results.} Light coupled to waveguides increases its temperature due to absorption and redistributes the impurities of the material, leading to thermo-optic and photorefractive effects, respectively, which change the effective refractive index of the waveguide. These effects are particularly stronger at shorter wavelengths. As the pump power increases, we observed a shift in the cavity resonance towards higher frequencies, as depicted in Figure~\ref{fig:classical}A. The direction of the shift indicates that the photorefractive process is responsible~\cite{sun_nonlinear_2017}. However, for 1\,mW of pump power, we observed a slight redshift which can be explained by the thermo-optic effect being slightly more substantial at this power. By ignoring the thermo-optic effect and thermal expansion, and applying a linear fit to the corresponding shifts in Fig.~\ref{fig:classical}B, we estimated the photorefraction coefficient to be $\mathrm{-3.97(12)\times 10^{-3} W^{-1}}$, the same order of magnitude as in~\cite{mondain_photorefractive_2020}. Moreover, the photorefraction alters the phase-matching temperature for optimum parametric-down conversion (PDC) gain, as observed in Fig.~\ref{fig:classical}C. At low pump power, the first two points indicate no changes in the phase-matching temperature relative to the temperature of the SHG, for which we neglected photorefraction as the powers involved were low. However, we consistently noticed an increase in the phase-matching temperature with the increase in power, which agrees with a negative change in the refraction index. After correcting the phase mismatch due to photorefraction, the simultaneous fit of (de-) amplification in Fig.~\ref{fig:classical}D led to a threshold power of 60.0$\pm$1.4\,mW.

\begin{figure}[htbp]
\centering
\includegraphics[width=.45\textwidth]{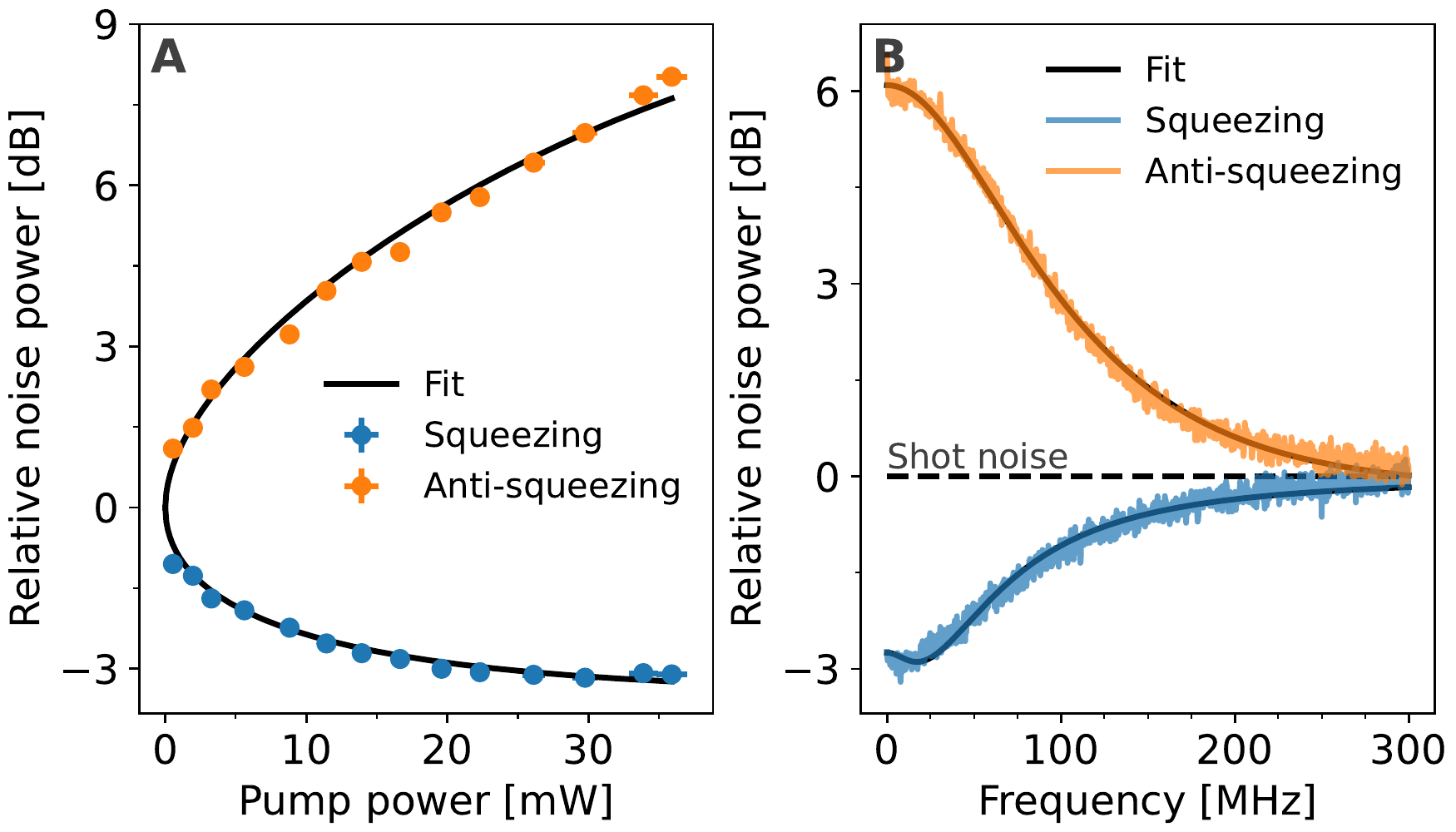} 
\caption{\label{fig:quantum} Quantum characterization of the Ti:PPLN squeezer. A) Anti-squeezing and squeezing levels at 5 MHz sideband frequency in terms of the pump power inside the nonlinear waveguide. B) Broadband squeezing generated by the resonator when pumped at 30 mW inside the waveguide.}
\end{figure}

Figure~\ref{fig:quantum}A shows the amount of detected squeezing and anti-squeezing at 5 MHz sideband frequency produced by the Ti:PPLN squeezer after optimizing the phase-matching. The maximum amount of detected anti-squeezing and squeezing were respectively $\mathrm{6.97(9)}$ and $\mathrm{-3.17(9)}$\,dB when the pump power inside the nonlinear waveguide was 30~mW.  The estimated total efficiency, including detection and escape efficiencies, is $\mathrm{62(4)}$\,\% from the simultaneous fit of (anti-) squeezing. Based on coatings specifications and internal propagation losses (10\,dB/m in a 12\,mm waveguide) we estimated the escape efficiency of the resonator to be 87\,\%. When considering the previously mentioned propagation and detection efficiency, the total efficiency from the fit is well explained. The lowest waveguide losses observed was 2\,dB/m~\cite{luo_direct_2015}, although not in the current batch. If these losses were achieved in the current waveguide, assuming all other experimental conditions were unchanged, 5\,dB of squeezing could be measured.

Figure~\ref{fig:quantum}B shows a spectrum of the squeezing and anti-squeezing noise power versus sideband frequency for a pump power of 30~mW. The (anti-) squeezing levels at low frequencies are compatible with the amount described in figure B. The full width half maximum (FWHM) obtained from the fit is 200 MHz, which is compatible with the estimation based on coatings and resonator length. Our model fitted to the data includes the effect of phase noise as this improved the quality of the fit~\cite{dwyer_squeezed_2013}, leading to an overall phase noise of the order of 20\,mrad.

\begin{figure}[htbp]
\centering
\includegraphics[width=.45\textwidth]{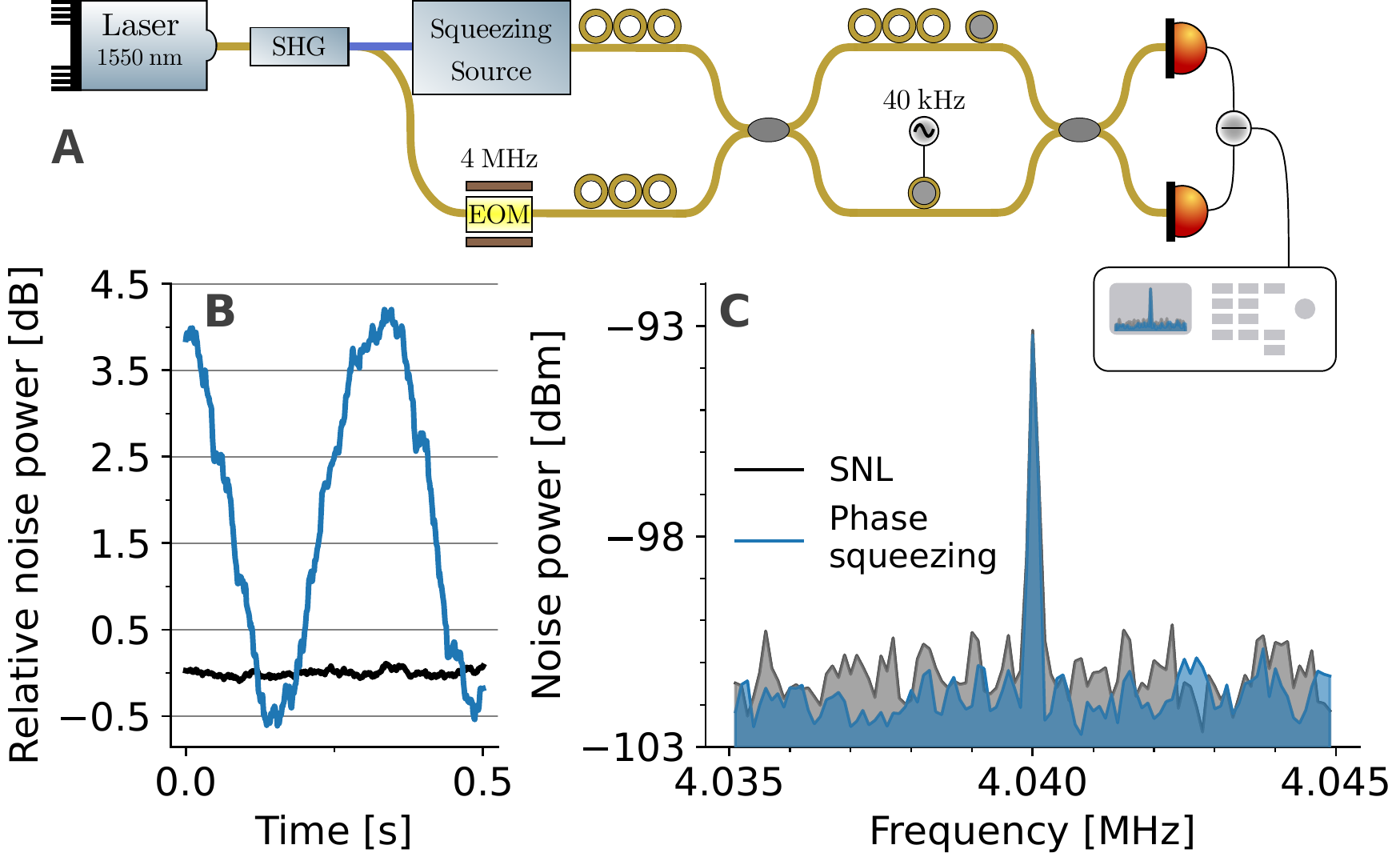} 
\caption{\label{fig:quantum sensing} Quantum sensing experiment. A) Setup for measuring phase noise at kHz range. B) Measurement of squeezing after the interferometer. Black and blue lines represent the normalized shot noise and squeezed vacuum noise, respectively. C) Quantum enhancement of the SNR when vacuum fluctuations (dark shading) or phase squeezing (blue shading) was injected into the interferometer.}
\end{figure}

\paragraph{Squeezed light enhanced phase sensing.} To demonstrate our fiber-coupled Ti:PPLN waveguide squeezer in a typical phase sensing experiment, we injected the squeezing into a Mach-Zehnder interferometer (MZI) as depicted in Fig.~\ref{fig:quantum sensing}A. The MZI was driven by a bright beam at 1550\,nm of 1\,mW power and locked at mid-fringe using a phase shifter (fiber stretcher) in one of the arms. The two outputs of the MZI were detected by photodiodes whose photocurrents were subtracted. A simulated phase signal inside one arm of the interferometer was generated at 40 kHz using another phase shifter in the other arm of the interferometer. To move away from classical noise, we intensity modulated the input beam of the MZI at 4\,MHz using an electro-optical modulator. The phase signal can then be observed as 40\,kHz sidebands of the 4 MHz modulation~\cite{liu_squeezing-enhanced_2017}. 

We first characterized the squeezed light in this setup without the amplitude modulation and the phase signal. Figure~\ref{fig:quantum sensing}B displays the noise power measured by the homodyne detector at a sideband frequency of 4.04\,MHz (resolution bandwidth 20\,kHz). We detected -0.5~dB of squeezing and 3.75\,dB anti-squeezing. For this measurement, we used a lower pump power to improve the stability of the system, where -2\,dB of squeezing was obtained right at the output of the squeezer. The reduction of squeezing is due to optical loss from the fiber components and the number of fiber connections. In particular, optimizing the MZI's visibility was challenging, and a maximum of 94\% was reached by carefully balancing the polarization and power inside the MZI. We expect that by using polarization-maintaining fibers instead, the visibility could be improved.

Figure~\ref{fig:quantum sensing}C shows the comparison of two measurements of the 40 kHz phase signal with and without squeezing. When no squeezing was injected into the unused port of the MZI, we observed an SNR of $\mathrm{8.26(25)~dB}$ (dark shading). With phase squeezing the observed SNR was $\mathrm{8.61(24)~dB}$ (blue shading). From the curves, it is clear that when using squeezing, the background noise is reduced while maintaining the same peak amplitude, leading to a quantum enhancement of the SNR.

\paragraph{Conclusion.} In conclusion, we presented an integrated source of single-mode squeezing guided by optical fibers. Using a fiber optical homodyne detector, we observed $\mathrm{-3.17(9)}$~dB of usable squeezing. We locked fiber alignments and optical phase to prepare the squeezed quadrature angle in homodyne detection. The system was robust for powers below 20\,mW, and squeezed light enhanced phase measurements were demonstrated, showing that our system is suitable for fiber-based sensing protocols. However, the system becomes inoperable for powers above 30\,mW. It is likely that this instability arises due to the onset of catastrophic photorefraction~\cite{carrascosa_understanding_2008}.

\begin{backmatter}
\bmsection{Funding} We acknowledge funding from the European Union’s Horizon 2020 research and innovation programme under grant agreement No 820474 (UNIQORN). RD, ULA, JSNN and TG acknowledge support by the Danish National Research Foundation, Center for Macroscopic Quantum States (bigQ, DNRF142).

\bmsection{Disclosures} The authors declare no conflicts of interest.
\end{backmatter}

\bibliography{references}

\bibliographyfullrefs{references}

\ifthenelse{\equal{\journalref}{aop}}{%
\section*{Author Biographies}
\begingroup
\setlength\intextsep{0pt}
\begin{minipage}[t][6.3cm][t]{1.0\textwidth} 
  \begin{wrapfigure}{L}{0.25\textwidth}
    \includegraphics[width=0.25\textwidth]{john_smith.eps}
  \end{wrapfigure}
  \noindent
  {\bfseries John Smith} received his BSc (Mathematics) in 2000 from The University of Maryland. His research interests include lasers and optics.
\end{minipage}
\begin{minipage}{1.0\textwidth}
  \begin{wrapfigure}{L}{0.25\textwidth}
    \includegraphics[width=0.25\textwidth]{alice_smith.eps}
  \end{wrapfigure}
  \noindent
  {\bfseries Alice Smith} also received her BSc (Mathematics) in 2000 from The University of Maryland. Her research interests also include lasers and optics.
\end{minipage}
\endgroup
}{}

\end{document}